\documentclass{article}
\usepackage{times}
\usepackage{amsfonts}
\usepackage{graphicx}
\begin{document}
\noindent
{\Large THE IRREVERSIBLE QUANTUM}
\vskip1cm
\noindent
{\bf P. Fern\'andez de C\'ordoba}$^{1,a}$, {\bf  J.M. Isidro}$^{1,b}$ {\bf Milton H. Perea}$^{1,2,c}$ and\\ {\bf J. Vazquez Molina}$^{1,d}$\\
${}^{1}$Instituto Universitario de Matem\'atica Pura y Aplicada,\\ Universidad Polit\'ecnica de Valencia, Valencia 46022, Spain\\
${}^{2}$Departamento de Matem\'aticas y F\'{\i}sica, Universidad Tecnol\'ogica\\ del Choc\'o, Colombia\\
${}^{a}${\tt pfernandez@mat.upv.es}, ${}^{b}${\tt joissan@mat.upv.es}\\ 
${}^{c}${\tt milpecr@posgrado.upv.es}, ${}^{d}${\tt joavzmo@etsii.upv.es}
\vskip.5cm
\vskip.5cm
\noindent
{\bf Abstract} 
We elaborate on the existing notion that quantum mechanics is an emergent phenomenon, by presenting a thermodynamical theory that is dual to quantum mechanics. This dual theory is that of classical irreversible thermodynamics. The linear regime of irreversibility considered here corresponds to the semiclassical approximation in quantum mechanics. An important issue we address is how the irreversibility of time evolution in thermodynamics is mapped onto the quantum--mechanical side of the correspondence. 


\section{Introduction}\label{uno}

In his Nobel Prize Lecture, Prigogine advocated an intriguing type of  ``complementarity between dynamics, which implies the knowledge of trajectories or wavefunctions, and thermodynamics, which implies entropy" \cite{PRIGOGINE1}. Another Nobel Prize winner, 't Hooft, has long argued that quantum mechanics must emerge from some underlying deterministic theory via information loss \cite{THOOFT3}. Entropy is of course intimately related to information loss, hence one expects some link to exist between these two approaches to quantum theory.

In an apparently unrelated venue, the  Chapman--Kolmogorov equation \cite{DOOB}
\begin{equation}
F(z_1)F(z_2)=F(z_1+z_2),
\label{funktional}
\end{equation}
is a functional equation in the unknown $F$, where $z_1, z_2$ are any two values assumed by the complex variable $z$. It has the general solution
\begin{equation}
F_a(z)={\rm e}^{za},
\label{loes}
\end{equation}
with $a\in\mathbb{C}$ an arbitrary constant. Implicitly assumed above is the multiplication rule for complex numbers. In other words, (\ref{loes}) solves  (\ref{funktional}) within a space of number--valued functions. If we allow for a more general multiplication rule such as matrix multiplication (possibly infinite--dimensional matrices), then the general solution (\ref{loes}) of the functional equation (\ref{funktional}) can be allowed to depend parametrically on a $z$--independent, constant {\it matrix}\/ or {\it operator}\/ $A$ acting on some linear space:
\begin{equation}
F_A(z)={\rm e}^{zA}.
\label{ssung}
\end{equation}
The functional equation (\ref{funktional}), in its different guises,  will play an important role in what follows. We see that its solutions are by no means unique, depending as they do on the space where one tries to solve the equation. Moreover, we will see that the question of specifying one solution space or another will bear a close relation to the question posed at the beginning---namely, the duality between thermodynamics and mechanics, on the one hand, and the emergence property of quantum mechanics, on the other.

Let ${\cal X}$ and ${\cal Y}$ respectively stand for the configuration spaces of a mechanical system and a thermodynamical system, the latter taken slightly away from equilibrium. We will be interested in the quantum theory based on ${\cal X}$, and in the theory of irreversible thermodynamics in the linear regime based on ${\cal Y}$ \cite{ONSAGER}. There exist profound analogies between these two theories \cite{NOI1, NOI2, OLAH, RUUGE1, RUUGE2}. Furthermore, seeming mismatches between the two actually have a natural explanation in the context of the emergent approach to quantum theory \cite{ADLER, CARROLL}; closely related topics were analysed long ago in \cite{DEBROGLIE} and more recently in \cite{ROVELLI, ELZE, FINSTER,  GALLEGO, PADDY, MATONE, PENROSE, CANADA, VELAZQUEZ}. One of these mismatches concerns the irreversibility of time evolution in the thermodynamical picture, as opposed to its reversibility in the quantum--mechanical picture.

The standard quantum formalism is invariant under time reversal. This is reflected, {\it e.g.}\/, in the fact that the Hilbert space of quantum states $L^2({\cal X})$ is complex and selfdual \cite{YOSIDA}, so one can exchange the incoming state $\vert\phi\rangle$ and the outgoing state $\langle\psi\vert$ by Hermitean conjugation, without ever stepping outside the given Hilbert space $L^2({\cal X})$. On the other hand, the thermodynamical space of states is the complex Banach space $L^1({\cal Y})$ of complex--valued, integrable probability densities $\phi:{\cal Y}\rightarrow\mathbb{C}$. This is in sharp contrast to the square--integrable probability density {\it amplitudes}\/ of quantum theory. Now the topological dual space to $L^1({\cal Y})$ is the Banach space $L^{\infty}({\cal Y})$ \cite{YOSIDA}. These two spaces fail to qualify as Hilbert spaces. In other words, for any $\vert\phi)\in L^1({\cal Y})$ and any $(\psi\vert\in L^{\infty}({\cal Y})$,\footnote{We follow the notations of ref. \cite{NOI1}. In particular, the round brackets in $\vert\phi)$ and $(\psi\vert$ refer to $L^1({\cal Y})$ and its topological dual $L^{\infty}({\cal Y})$, respectively, while the angular brackets of the quantum--mechanical ket $\vert \phi\rangle$ and bra $\langle \psi\vert$ refer to $L^2({\cal X})$ and its topological dual $L^2({\cal X})$. Concerning the measure on ${\cal X}$ and ${\cal Y}$, see below.} the respective norms $\vert\vert\phi\vert\vert_1$ and $\vert\vert\psi\vert\vert_{\infty}$ are well defined, but neither of these derives from a scalar product. All there exists is a nondegenerate, bilinear pairing
\begin{equation}
(\,\cdot\vert\cdot\,):L^{\infty}({\cal Y})\times L^{1}({\cal Y})\longrightarrow\mathbb{C}
\label{banaj}
\end{equation}
taking the covector $(\psi\vert$ and the vector $\vert\phi)$ into the number $(\psi\vert\phi)$:
\begin{equation}
(\psi\vert\phi):=\int_{\cal Y}\psi^*\phi.
\label{gungen}
\end{equation}
Under these circumstances there is no exchanging the incoming state $\vert\phi)\in L^1({\cal Y})$ and the outgoing state $(\psi\vert\in L^{\infty}({\cal Y})$, as they belong to different spaces.  Therefore time reversal symmetry is lost. We see that {\it dispensing with the scalar product in quantum theory is the same as dispensing with time reversal symmetry}\/.  

We have in \cite{NOI1, NOI2} tocuhed on several basic issues concerning a thermodynamical formalism for quantum theory. Specifically, a map has been constructed between the quantum mechanics of a finite number of degrees of freedom, on the one hand, and the theory of irreversible processes in the linear regime, on the other. The current paper elaborates further on the properties of a {\it thermodynamical dual theory}\/ for emergent quantum mechanics. The underlying logic might be briefly summarised as follows:\\
{\it i)}\/ it has been claimed that thermodynamics is complementary, or dual, to mechanics;\\
{\it ii)}\/ mechanics is symmetric under time reversal while thermodynamics is not;\\
{\it iii)}\/ dispensing with time reversal symmetry is the same as dispensing with the scalar product in quantum theory;\\  
{\it iv)}\/ the representation of the Chapman--Kolmogorov equation (\ref{funktional}) on the quantum mechanical Hilbert space $L^2({\cal X})$ makes decisive use of the scalar product;\\
{\it v)}\/ here we construct representations of (\ref{funktional}) on the thermodynamical Banach spaces $L^1({\cal Y})$ and $L^{\infty}({\cal Y})$, where no scalar product is present.

{}For simplicity we will henceforth assume ${\cal X}$ and ${\cal Y}$ both equal to $\mathbb{R}$, the latter endowed with the Lebesgue measure.

The aim of our paper is not to reformulate the theory of irreversible thermodynamics as originally developed in \cite{ONSAGER}. Rather, {\it we intend to exhibit irreversibility as a key property of quantum--mechanical behaviour}\/.

\section{Different representations for Chapman--Kolmogorov}

\subsection{The quantum--mechanical representation}

In quantum mechanics it is customary to write (\ref{funktional}) as
\begin{equation}
U(t_1)U(t_2)=U(t_1+t_2), \qquad t\in\mathbb{R},
\label{muchauuu}
\end{equation}
and to call it the {\it group property}\/ of time evolution. If $H$ denotes the quantum Hamiltonian operator (assumed time--independent for simplicity), then (\ref{muchauuu}) is solved by matrices such as (\ref{ssung}), here called time--evolution operators and defined as
\begin{equation}
U(t):=\exp\left(-\frac{{\rm i}}{\hbar}t{H}\right).
\label{uniprop}
\end{equation}
The solutions of (\ref{muchauuu}) satisfy the differential equation
\begin{equation}
{\rm i}\hbar\frac{{\rm d}U}{{\rm d}t}=HU(t),\qquad H={\rm i}\hbar\frac{{\rm d}U}{{\rm d}t}{\Big \vert}_{t=0}.
\label{ouh}
\end{equation}
Comparing (\ref{uniprop}) with (\ref{ssung}) we have $z=t$ and $A=-{\rm i}H/\hbar$. The $U(t)$ are unitary on $L^2(\mathbb{R})$. In a basis of position eigenfunctions $\vert x\rangle$, the matrix elements of $U(t)$ equal the Feynman propagator: $\langle x_2\vert  U(t_2-t_1)\vert x_1\rangle=K\left(x_2,t_2\vert x_1,t_1\right)$. In terms of the latter, one rewrites the group property (\ref{muchauuu}) as
\begin{equation}
K\left(x_3,t_3\vert x_1,t_1\right)=\int{\rm d}x_2\,K\left(x_3,t_3\vert x_2,t_2\right)K\left(x_2,t_2\vert x_1,t_1\right).
\label{gruppo}
\end{equation}
There is a path integral for the Feynman propagator $K$:
\begin{equation}
K\left(x_2,t_2\vert x_1,t_1\right)=\int_{x(t_1)=x_1}^{x(t_2)=x_2}{\rm D}x(t)\,\exp\left\{\frac{{\rm i}}{\hbar}\int_{t_1}^{t_2}{\rm d}t\,L\left[x(t),\dot x(t)\right]\right\},
\label{paz}
\end{equation}
where $L$ is the classical Lagrangian function. 

To summarise, the operators (\ref{uniprop}) provide a unitary representation of the commutative group (\ref{muchauuu}) on the Hilbert space $L^2(\mathbb{R})$.

\subsection{Intermezzo}\label{intermdd}

Here we recall some technicalities to be used later; a good general reference is \cite{YOSIDA}.

$L^1(\mathbb{R})$ is the space of all Lebesgue measurable, absolutely integrable functions $\phi:\mathbb{R}\longrightarrow\mathbb{C}$, {\it i.e.}\/, functions such that $\int_{\mathbb{R}}\vert \phi(y)\vert{\rm d}y<\infty$. This is a complex Banach space with respect to the norm $\vert\vert \psi\vert\vert_1:=\int_{\mathbb{R}}\vert \phi(y)\vert{\rm d}y$.\footnote{Just for comparison, the norm on the Hilbert space $L^2(\mathbb{R})$ is $\vert\vert \phi\vert\vert_2:=\left(\int_{\mathbb{R}}\vert \phi(y)\vert^2{\rm d}y\right)^{1/2}$.} A a {\it denumerable}\/ basis (a Schauder basis) exists for $L^1(\mathbb{R})$.

The topological dual space to $L^1(\mathbb{R})$ is $L^{\infty}(\mathbb{R})$, a duality between the two being given in Eqs. (\ref{banaj}), (\ref{gungen}). $L^{\infty}(\mathbb{R})$ is the space of all Lebesgue measurable functions $\psi:\mathbb{R}\longrightarrow\mathbb{C}$ that are essentially bounded, {\it i.e.}\/, functions that remain bounded on all $\mathbb{R}$ except possibly on a set of measure zero. $L^{\infty}(\mathbb{R})$ is a Banach space with respect to the norm $\vert\vert \cdot\vert\vert_{\infty}$, defined as follows. A nonnegative number $\alpha\in\mathbb{R}$ is said to be an essential upper bound of $\psi$ whenever the set of points $y\in \mathbb{R}$ where $\vert\psi(y)\vert\geq \alpha$ has zero measure. The norm $\vert\vert\psi\vert\vert_{\infty}$ is the infimum of all those $\alpha$:
\begin{equation}
\vert\vert\psi\vert\vert_{\infty}:={\rm inf}\left\{\alpha\in\mathbb{R}^+: \alpha\;{\rm  essential}\; {\rm upper}\; {\rm bound}\; {\rm of}\;\psi\right\}.
\label{eseninf}
\end{equation}
A key property is that one can pointwise multiply $\psi\in L^{\infty}(\mathbb{R})$ with $\phi\in L^1(\mathbb{R})$ to obtain $\psi\phi\in L^1(\mathbb{R})$ because $\int_{\mathbb{R}} \vert \psi\phi\vert{\rm d}y<\infty$; this is used decisively in the pairing (\ref{gungen}). Another key property of $L^{\infty}(\mathbb{R})$ is that it admits no Schauder basis.

The space $L^1(\mathbb{R})$ is canonically and isometrically embedded into its topological bidual, {\it i.e.}\/, $L^1(\mathbb{R})\subset L^1(\mathbb{R})^{**}$. Since $L^1(\mathbb{R})$ is nonreflexive, this inclusion is strict, a property that will be used later on.\footnote{The topological complementary space to $L^1(\mathbb{R})$, {\it i.e.}\/, the space $Z$ such that $L^1(\mathbb{R})^{**}=L^1(\mathbb{R})\oplus Z$, is known in the literature, but it will not be necessary here.} Finally, the absence of a scalar product on $L^1(\mathbb{R})$ and $L^{\infty}(\mathbb{R})$ does not prevent the existence of unitary operators on them, the latter being defined as those that preserve the corresponding norm.

\subsection{The representation in irreversible thermodynamics}

In statistics, the Chapman--Kolmogorov equation (\ref{funktional}) was well known before the advent of quantum theory \cite{DOOB}. Here one is given a certain measure space ${\cal Y}$ (here assumed equal to $\mathbb{R}$ endowed with the Lebesgue measure) and the corresponding Banach spaces $L^1(\mathbb{R})$ and its topological dual $L^{\infty}(\mathbb{R})$.  These two will become carrier spaces for representations of the Chapman--Kolmogorov equation (\ref{funktional}).

One calls $f_1\left({y_{2}\atop \tau_{2}}{\Big\vert}{y_1\atop \tau_1}\right)$ the {\it conditional}\/ probability that the random variable $y\in \mathbb{R}$ takes on the value $y_2$ at time $\tau_2$ provided that it took on the value $y_1$ at time $\tau_1$. Then one usually writes the Chapman--Kolmogorov equation (\ref{funktional}) in a manner similar to (\ref{gruppo}),
\begin{equation}
f_1\left({y_{3}\atop \tau_{3}}{\Big\vert}{y_1\atop \tau_1}\right)=\int{\rm d}y_2\, f_1\left({y_{3}\atop \tau_{3}}{\Big\vert}{y_2\atop \tau_2}\right)f_1\left({y_{2}\atop \tau_{2}}{\Big\vert}{y_1\atop \tau_1}\right),
\label{jjaev}
\end{equation}
which expresses the Bayes rule for conditional probabilities. A representation of this equation by means of linear operators ${\cal U}(\tau)$ on $L^1(\mathbb{R})$ and on $L^{\infty}(\mathbb{R})$ would thus have to satisfy the algebra
\begin{equation}
{\cal U}(\tau_1)\,{\cal U}(\tau_2)={\cal U}(\tau_1+\tau_2),
\label{ohlala}
\end{equation}
which is again a presentation of (\ref{funktional}). We can immediately read off the matrix elements of ${\cal U}(\tau)$:
\begin{equation}
(y_2\vert {\cal U}(\tau_2-\tau_1)\vert y_1)=f_1\left({y_2\atop\tau_2}{\Big\vert}{y_1\atop\tau_1}\right).
\label{inkons}
\end{equation}
As opposed to the quantum--mechanical case, the carrier space for the representation of the algebra (\ref{ohlala}) is Banach but not Hilbert. The reason for this is that one deals directly with probabilities rather than amplitudes.

The question arises: if one were to express the matrix (\ref{inkons}) in the form given by the general solution (\ref{ssung}), then clearly one would have $z=\tau$, but what would the operator $A$ be? It is mathematically true, though physically unsatisfactory, to claim that $A$ would be (proportional to) the logarithm of ${\cal U}(\tau)$.
One of the purposes of this paper is to determine the operator $A$ explicitly, and to interpret it in the terms stated in the introduction. However, in order to do this, a knowledge of the conditional probabilities $f_1\left({y_{2}\atop \tau_{2}}{\Big\vert}{y_1\atop \tau_1}\right)$ is needed.

There are a number of instances in which the $f_1\left({y_{2}\atop \tau_{2}}{\Big\vert}{y_1\atop \tau_1}\right)$  are known explicitly. An important example is that of {\it classical, irreversible thermodynamics of stationary, Markov processes in the linear regime}\/. For such processes one has \cite{ONSAGER}
\begin{equation}
f_1\left({y_{2}\atop \tau_2}{\Big\vert}{y_{1}\atop \tau_1}\right)
=\frac{1}{\sqrt{2\pi}}\frac{s/k_B}{\sqrt{1-{\rm e}^{-2\gamma(\tau_2-\tau_1)}}}
\exp\left[-\frac{s}{2k_B}\frac{\left(y_{2}-{\rm e}^{-\gamma(\tau_2-\tau_1)}y_{1}\right)^2}{1-{\rm e}^{-2\gamma(\tau_2-\tau_1)}}\right].
\label{proba}
\end{equation}
The notation used here is that of \cite{NOI1}. Specifically, $k_B$ is Boltzmann's constant, the entropy $S$ is a function of the extensive parameter $y$, and we expand $S$ in a Taylor series around a stable equilibrium point. Up to quadratic terms we have
\begin{equation}
S=S_0-\frac{1}{2}sy^2+\ldots, \qquad s:=-\frac{{\rm d}^2 S}{{\rm d} y^2}\Big\vert_{0}>0.
\label{teilor}
\end{equation}
Moreover, the assumption of linearity implies the following proportionality between the thermodynamical force $Y:={\rm d}S/{\rm d}y$ and the flux $\dot y:={\rm d}y/{\rm d}\tau$ it produces \cite{ONSAGER}:
\begin{equation}
\dot y=L Y, \qquad L>0.
\label{eleerre}
\end{equation}
The Onsager coefficient $L$ must be positive for the process to be dissipative. Finally $\gamma:=sL$. Sometimes one also uses $R:=L^{-1}$, so $\gamma=s/R$.

The following path--integral representation for the conditional probabilities (\ref{proba}) of these models is noteworthy \cite{ONSAGER}:
\begin{equation}
f_1\left({y_{2}\atop \tau_{2}}{\Big\vert}{y_1\atop \tau_1}\right)=\int_{y(\tau_1)=y_1}^{y(\tau_2)=y_2}{\rm D}y(\tau)\,
\exp\left\{-\frac{1}{2k_B}\int_{\tau_1}^{\tau_{2}}{\rm d}\tau\,{\cal L}\left[\dot y(\tau), y(\tau)\right]\right\}.
\label{cepele}
\end{equation}
The  above exponential contains the {\it thermodynamical Lagrangian}\/ ${\cal L}$, defined as
\begin{equation}
{\cal L}\left[\dot y(\tau), y(\tau)\right]:=\frac{R}{2}\left[\dot y^2(\tau)+\gamma^2y^2(\tau)\right], \quad \dot y:=\frac{{\rm d}y}{{\rm d}\tau}.
\label{agra}
\end{equation}
The path integral (\ref{cepele}) is the thermodynamical analogue of (\ref{paz}). The corresponding thermodynamical momentum $p_y$ equals $R{\rm d}y/{\rm d}\tau$,
where $R$ plays the role of a mass, and the {\it thermodynamical Hamiltonian}\/ ${\cal H}$ corresponding to (\ref{agra}) reads
\begin{equation}
{\cal H}=\frac{1}{2R}p_y^2-\frac{R\gamma^2}{2}y^2.
\label{jjamm}
\end{equation}
It must be borne in mind, however, that the dimensions of ${\cal L}$ and ${\cal H}$ are entropy per unit time. With this caveat, we will continue to call ${\cal H}$ a Hamiltonian.

\subsection{Mapping irreversible thermodynamics into quantum mechanics}\label{siete}

{}For the processes considered in (\ref{proba}) we claim that one can define operators on $L^1(\mathbb{R})$ and on $L^{\infty}(\mathbb{R})$
\begin{equation}
{\cal U}(\tau):=\exp\left(-\frac{1}{2k_B}\tau {\cal H}\right)
\label{juguete}
\end{equation}
with ${\cal H}$ suitably chosen, such that their matrix elements coincide with those given in (\ref{inkons}). Hence the ${\cal U}(\tau)$ will provide a representation of the algebra (\ref{ohlala}). In what follows we construct ${\cal U}(\tau)$ explicitly, but one can already expect the argument ${\cal H}$ of the exponential (\ref{juguete}) to be some {\it operator}\/ version of the thermodynamical Hamiltonian {\it function}\/ given in (\ref{jjamm}). For this reason we have not distinguished notationally between the two. This operator ${\cal H}$ will also turn out to be (proportional to) the unknown operator $A$ mentioned after eq. (\ref{inkons}).
From (\ref{juguete}) it follows that the thermodynamical analogue of the quantum--mechanical equation (\ref{ouh}) is
\begin{equation}
-2k_B\frac{{\rm d}\,{\cal U}(\tau)}{{\rm d}\tau}={\cal H}\,{\cal U}(\tau), \qquad {\cal H}=-2k_B\frac{{\rm d}\,{\cal U}(\tau)}{{\rm d}\tau}{\Big\vert}_{\tau=0}.
\label{magister}
\end{equation} 

We can resort to our previous work \cite{NOI1} in order to identify the operator ${\cal H}$ in its action on $L^1(\mathbb{R})$ and on $L^{\infty}(\mathbb{R})$. In \cite{NOI1} we have established a map between quantum mechanics in the semiclassical regime, on the one hand, and the theory of classical, irreversible thermodynamics of stationary, Markov processes in the linear regime, on the other hand. In the mechanical picture, the relevant Lagrangian and Hamiltonian functions are
\begin{equation}
L=\frac{m}{2}\left(\frac{{\rm d}x}{{\rm d}t}\right)^2-\frac{m\omega^2}{2}x^2, \qquad H=\frac{1}{2m}p_x^2+\frac{m\omega^2}{2}x^2.
\label{mecalang}
\end{equation}
Comparing them with their thermodynamical partners (\ref{agra}) and (\ref{jjamm}), we see that the mechanical and the thermodynamical functions can be transformed
into each other if we apply the replacements\footnote{While the first two replacements of (\ref{bazz}) are dimensionally correct without any further assumptions, the third identification also requires that $x$ and $y$ have the same dimensions. Since this need not always be the case, a dimensionful conversion factor must be understood as implicitly contained in the replacement $x\leftrightarrow y$, whenever needed.}
\begin{equation}
\omega\leftrightarrow\gamma,\qquad \frac{m\omega}{\hbar}\leftrightarrow\frac{s}{2k_B}, \qquad x\leftrightarrow y,
\label{bazz}
\end{equation}
as well as the Wick rotation
\begin{equation}
\tau={\rm i}t.
\label{vic}
\end{equation}
Furthermore, Boltzmann's constant $k_B$ is the thermodynamical partner of Planck's constant $\hbar$ multiplied by 2 \cite{RUUGE1}:
\begin{equation}
\hbar\leftrightarrow 2k_B.
\label{komp}
\end{equation}
As a consistency check one can apply all the above replacements to (\ref{uniprop}) in order to arrive at
\begin{equation}
U(t)=\exp\left(-\frac{{\rm i}}{\hbar}tH\right) \leftrightarrow\exp\left(-\frac{1}{2k_B}\tau {\cal H}\right)={\cal U}(\tau).
\label{aniu}
\end{equation}
However, we still have to identify the operator ${\cal H}$ in its action on thermodynamical states. This will be done in section \ref{banij}.

\subsection{Incoming states {\it vs}\/. outgoing states}

In principle, thermodynamical states are normalised probability densities, hence elements of $L^1(\mathbb{R})$. However, as we will see shortly, this viewpoint must be extended somewhat. For this purpose let us call the elements of $L^1(\mathbb{R})$ {\it incoming states}\/. Incoming linear operators ${\cal O}_{\rm in}$ are defined
\begin{equation}
{\cal O}_{\rm in}:L^1(\mathbb{R})\longrightarrow L^1(\mathbb{R}),
\label{iniziale}
\end{equation}
so as to map incoming states $\vert\phi)\in L^1(\mathbb{R})$ into incoming states ${\cal O}_{\rm in}\vert\phi)\in L^1(\mathbb{R})$. Incoming states are postulated to evolve in time according to
\begin{equation}
-2k_B\frac{{\rm d}\vert\phi)}{{\rm d}\tau}={\cal H}_{\rm in}\vert\phi),
\label{calor}
\end{equation}
where ${\cal H}_{\rm in}$ is an incoming linear operator, to be identified presently. 

The space of outgoing states is the topological dual of $L^1(\mathbb{R})$, hence $L^{\infty}(\mathbb{R})$. Outgoing linear operators ${\cal O}_{\rm out}$ are similarly defined
\begin{equation}
{\cal O}_{\rm out}:L^{\infty}(\mathbb{R})\longrightarrow L^{\infty}(\mathbb{R}),
\label{finale}
\end{equation}
in order to map outgoing states $(\psi\vert\in L^{\infty}(\mathbb{R})$ into outgoing states $(\psi\vert{\cal O}_{\rm out}\in L^{\infty}(\mathbb{R})$. 
The operator ${\cal O}^{T}_{\rm in}$ that is transpose to an incoming operator ${\cal O}_{\rm in}$ is defined on the topological dual space:
\begin{equation}
{\cal O}^{T}_{\rm in}:L^{\infty}(\mathbb{R})\longrightarrow L^{\infty}(\mathbb{R}).
\label{duale}
\end{equation}
In this way ${\cal O}^{T}_{\rm in}$ is actually an outgoing operator ${\cal O}_{\rm out}$.\footnote{Since the topological bidual $(L^1(\mathbb{R}))^{**}$ contains more than just $L^{1}(\mathbb{R})$, we stop short of stating that ``The transpose ${\cal O}^{T}_{\rm out}$ to an outgoing operator ${\cal O}_{\rm out}$ is an incoming operator ${\cal O}_{\rm in}$". The previous statement, trivially true in finitely many dimensions and still true on $L^2(\mathbb{R})$, no longer holds in our context, with the consequence that twice transposing does not give back the original operator. We will see in section \ref{spek} that this fact has far--reaching implications.} By definition the transpose satisfies
\begin{equation}
(\psi\vert {\cal O}_{\rm in}^{T}\vert\phi)=(\psi\vert{\cal O}_{\rm in}\vert\phi), \qquad\forall\, (\psi\vert\in L^{\infty}(\mathbb{R}),\, \forall\,\vert\phi)\in L^1(\mathbb{R}).
\label{satisfacendo}
\end{equation} 

What equation should govern the time evolution of outgoing states? Clearly it can only be
\begin{equation}
-2k_B\frac{{\rm d}(\psi\vert}{{\rm d}\tau}=(\psi\vert{\cal H}^{T}_{\rm in}=(\psi\vert{\cal H}_{\rm out},
\label{claramente}
\end{equation}
therefore
\begin{equation}
-2k_B\frac{{\rm d}}{{\rm d}\tau}(\psi\vert\phi)=(\psi\vert{\cal H}_{\rm in}^{T}\vert\phi)+(\psi\vert{\cal H}_{\rm in}\vert\phi).
\label{cott}
\end{equation}
The right--hand side of the above is generally nonzero: it expresses the irreversibility property of time evolution in thermodynamics.  This is a far cry from the time--symmetric case of standard quantum  mechanics, where ${\rm i}\hbar{\rm d}(\langle\psi\vert\phi\rangle)/{\rm d}t=0$.

One further point deserves attention. In standard quantum mechanics on $L^2(\mathbb{R})$, the matrix element $\langle\psi\vert{\cal O}\vert\phi\rangle =\int{\rm d}x\,\psi^*(x){\cal O}\phi(x)$ naturally carries the dimensions of the operator ${\cal O}$; here both $\psi^*(x)$ and $\phi(x)$ have the dimension $[x]^{-1/2}$ of a probability amplitude on $\mathbb{R}$. In the thermodynamical dual to quantum theory, the incoming state $\vert\phi)\in L^1(\mathbb{R})$ carries the dimension $[y]^{-1}$ because it is a probability {\it density}\/, while the outgoing state $(\psi\vert\in L^{\infty}(\mathbb{R})$ is {\it dimensionless}\/ because it is {\it not}\/ meant to be integrated on its own. It is only upon taking the pairing (\ref{gungen}) that $(\psi\vert$ will be integrated against ${\cal O}\vert\phi)$. So the dimensions of $(\psi\vert{\cal O}\vert\phi)$ are again correct, although the dimensional balance between incoming and outgoing states that existed in $L^2(\mathbb{R})$ has disappeared. 

Altogether, dispensing with the scalar product in quantum theory is the same as dispensing with time reversal symmetry. Moreover, dispensing with the scalar product has the consequence that, as thermodynamical states, one must regard not just the elements of $L^1(\mathbb{R})$ but also those of its topological dual $L^{\infty}(\mathbb{R})$.

\section{The harmonic oscillator representation of irreversible thermodynamics}

{}For mechanics we use the {\it dimensionless}\/ coordinate $x\in\mathbb{R}$. Then the quantum harmonic oscillator equation on $L^2(\mathbb{R})$ reads 
\begin{equation}
\left(-\frac{{\rm d}^2}{{\rm d}x^2}+x^2\right)w(x)=\varepsilon w(x), \qquad \varepsilon\in\mathbb{R},
\label{sabrosura}
\end{equation}
where $\varepsilon$ is a dimensionless energy eigenvalue.

\subsection{The oscillator on the Banach spaces $L^1(\mathbb{R})$ and $L^{\infty}(\mathbb{R})$}\label{banij}

{}For thermodynamics we  use the {\it dimensionless}\/ coordinate $y\in\mathbb{R}$. Then the dimensionless thermodynamical momentum is represented as $-{\rm i}{\rm d}/{\rm d}y$, and the equation for the thermodynamical oscillator reads
\begin{equation}
-\left(\frac{{\rm d}^2}{{\rm d}y^2}+y^2\right)w(y)=\sigma w(y)\qquad \sigma\in\mathbb{R}.
\label{auto}
\end{equation}
Above, $\sigma$ is a dimensionless eigenvalue (entropy per unit time), which we require to be real for physical reasons. 
With respect to (\ref{sabrosura}), the only change in (\ref{auto}) is the sign of the potential term (see (\ref{agra}) and (\ref{jjamm})). Eq. (\ref{auto}) identifies the operator ${\cal H}$ explicitly in its action on $L^1(\mathbb{R})$ and $L^{\infty}(\mathbb{R})$, a question posed in section \ref{siete}. Specifically, for the action of the Hamiltonian on the initial states we have
\begin{equation}
{\cal H}_{\rm in}=-\frac{{\rm d}^2}{{\rm d}y^2}-y^2:L^{1}(\mathbb{R})\longrightarrow L^{1}(\mathbb{R}).
\label{miglia}
\end{equation}
The operator ${\cal H}_{\rm out}$ is formally the same as ${\cal H}_{\rm in}$, but it acts on the dual space:
\begin{equation}
{\cal H}_{\rm out}=-\frac{{\rm d}^2}{{\rm d}y^2}-y^2:L^{\infty}(\mathbb{R})\longrightarrow L^{\infty}(\mathbb{R}).
\label{venti}
\end{equation}

In order to solve (\ref{auto}) we first look for a factorisation of $w(y)$ in the form
\begin{equation}
w(y)=h(y)\exp(\alpha y^2),\qquad \alpha\in\mathbb{C},
\label{fakto}
\end{equation}
where $\alpha$ is some constant to be picked appropriately. With (\ref{fakto}) in (\ref{auto}) one finds
\begin{equation}
\frac{{\rm d}^2}{{\rm d}y^2}h(y)+4\alpha y \frac{{\rm d}}{{\rm d}y}h(y)+\left[(2\alpha+\sigma)+(4\alpha^2+1)y^2\right]h(y)=0.
\label{rizat}
\end{equation}
The choice $\alpha={\rm i}/2$ simplifies (\ref{rizat}) considerably:
\begin{equation}
\frac{{\rm d}^2}{{\rm d}y^2}h(y)+ 2{\rm i}y \frac{{\rm d}}{{\rm d}y}h(y)+({\rm i}+\sigma)h(y)=0.
\label{facil}
\end{equation}
{}Finally the change of variables $z={\rm e}^{{\rm i}\frac{3\pi}{4}}y$ reduces (\ref{facil}) to 
\begin{equation}
\frac{{\rm d}^2}{{\rm d}z^2}\tilde h(z)-2z\frac{{\rm d}}{{\rm d}z}\tilde h(z)-(1-{\rm i}\sigma)\tilde h(z)=0,
\label{ident}
\end{equation}
where we have defined $\tilde h(z):=h\left({\rm e}^{-{\rm i}\frac{3\pi}{4}}z\right)=h(y)$. Now (\ref{ident}) is a particular instance of the Hermite differential equation on the complex plane, 
\begin{equation}
H''(z)-2zH'(z)+2\nu H(z)=0, \qquad \nu\in\mathbb{C}.
\label{ermitico}
\end{equation}
In our case we have $2\nu=-1+{\rm i}\sigma$ with $\sigma\in\mathbb{R}$, so $\nu\notin\mathbb{N}$. When $\nu\notin\mathbb{N}$ two linearly independent solutions to the Hermite equation are given by the Hermite functions $H_{\nu}(z)$ and $H_{\nu}(-z)$, where \cite{LEBEDEV}
\begin{equation}
H_{\nu}(z)=\frac{1}{2\Gamma(-\nu)}\sum_{n=0}^{\infty}\frac{(-1)^n\Gamma\left(\frac{n-\nu}{2}\right)}{n!}(2z)^{n}.
\label{melhor}
\end{equation}
The above power series defines an entire function of $z\in\mathbb{C}$ for any value of $\nu\in\mathbb{C}$. Its asymptotic behaviour is \cite{LEBEDEV}:
\begin{equation}
H_{\nu}(z)\sim (2z)^{\nu}-\frac{\sqrt{\pi}{\rm e}^{{\rm i}\pi\nu}}{\Gamma(-\nu)}\,z^{-\nu-1}\,{\rm e}^{z^2},\quad 
\vert z\vert\to\infty,\quad\pi/4<\arg(z)<5\pi/4.
\label{zigg}
\end{equation}
In (\ref{zigg}) we have dropped subdominant terms, keeping only the leading contributions; the angular sector $\quad\pi/4<\arg(z)<5\pi/4$ is imposed on us by the change of variables $z={\rm e}^{{\rm i}\frac{3\pi}{4}}y$ made above for $y\in\mathbb{R}$.

Altogether, two linearly independent solutions to (\ref{auto}) corresponding to the eigenvalue $\sigma\in\mathbb{R}$ are given by $w^{\pm}_{\sigma}(y)$, where
\begin{equation}
w^{\pm}_{\sigma}(y):=H_{-\frac{1}{2}+\frac{{\rm i}\sigma}{2}}\left(\pm{\rm e}^{{\rm i}\frac{3\pi}{4}}y\right){\rm e}^{{\rm i}y^2/2}.
\label{aniv}
\end{equation}
By (\ref{zigg}), their asymptotic behaviour for $\vert y\vert\to\infty$ is
\begin{equation}
w^{\pm}_{\sigma}(y)\sim\left(\pm2{\rm e}^{{\rm i}\frac{3\pi}{4}}y\right)^{-\frac{1}{2}+\frac{{\rm i}\sigma}{2}}{\rm e}^{{\rm i}y^2/2}
-\frac{\sqrt{\pi}{\rm e}^{-\pi(\sigma+{\rm i})/2}}{\Gamma\left(\frac{1-{\rm i}\sigma}{2}\right)}\,\left(\pm{\rm e}^{{\rm i}\frac{3\pi}{4}}y\right)^{-\frac{1}{2}-\frac{{\rm i}\sigma}{2}}\,{\rm e}^{-{\rm i}y^2/2}.
\label{asstt}
\end{equation}
We are looking for eigenfunctions within $L^1(\mathbb{R})$ and/or $L^{\infty}(\mathbb{R})$. Eqn. (\ref{asstt}) proves that $w^{\pm}_{\sigma}(y)\in L^{\infty}(\mathbb{R})$ but $w^{\pm}_{\sigma}(y)\notin L^1(\mathbb{R})$.

\subsection{The spectrum}\label{spek}

Summarising,  the operator $-{\rm d}^2/{\rm d}y^2-y^2$ on $L^{\infty}(\mathbb{R})$ has an eigenvalue spectrum containing the whole real line $\mathbb{R}$.\footnote{Actually the eigenvalue spectrum of this operator on $L^{\infty}(\mathbb{R})$ also contains nonreal eigenvalues (see (\ref{dekret})), but here we are only interested in real eigenvalues.} This spectrum is twice degenerate, the (unnormalised) eigenfunctions corresponding to $\sigma\in\mathbb{R}$ being given in Eq. (\ref{aniv}). The same operator acting on $L^1(\mathbb{R})$ has a void spectrum. This latter conclusion is not as tragic as it might seem at first sight---on the contrary, everything fits together once one realises that evolution in thermodynamical time $\tau$ is irreversible, and that the space $L^1(\mathbb{R})$, which admits a Schauder basis, has a topological  dual $L^{\infty}(\mathbb{R})$ admitting no Schauder basis. Let us analyse these facts from a physical and from a mathematical viewpoint.

Physically, an empty spectrum on $L^1(\mathbb{R})$ just means that {\it there can be no incoming eigenstates}\/. Moreover, no incoming state can ever evolve into an incoming {\it eigenstate}\/ under thermodynamical evolution. This is an expression of irreversibility. However, as a result of evolution in $\tau$, one can perfectly well obtain {\it outgoing}\/ eigenstates. The latter remain outgoing {\it eigenstates}\/ under thermodynamical evolution. 

Mathematically, in standard quantum mechanics on $L^2(\mathbb{R})$ one is used to taking the transpose of a matrix by exchanging rows with columns. Implicitly understood here is the existence of Schauder bases in the space of $L^2(\mathbb{R})$ and in its topological dual (again $L^2(\mathbb{R})$). Once one diagonalises an operator, how can it be that its transpose is not diagonal as well? While this cannot happen in $L^2(\mathbb{R})$, {\it this can perfectly well be the case when dealing with the spaces $L^1(\mathbb{R})$ and $L^{\infty}(\mathbb{R})$, because $L^1(\mathbb{R})$ admits a Schauder basis while $L^{\infty}(\mathbb{R})$ does not}\/.
In turn, this is a consequence of the fact that we are renouncing probability density {\it amplitudes}\/ (elements of $L^2(\mathbb{R})$) in favour of probability {\it densities}\/ (elements of $L^1(\mathbb{R})$), as befits a thermodynamical description of quantum theory. 

One would like to identify the thermodynamical analogue of the quantum mechanical vacuum state; one expects to somehow map the quantum--mechanical state of least energy, or vacuum, into the thermodynamical state of maximal entropy. Let us recall that the (unnormalised) quantum--mechanical vacuum wavefunction is $\exp(-x^2/2)$. The Wick rotation (\ref{vic}) introduces the imaginary unit, giving us the term $\exp({\rm i}y^2/2)$ in (\ref{aniv}). Now  $\nu=-1/2+{\rm i}\sigma/2=0$ only when $\sigma=-{\rm i}$, a possibility we have excluded per decree. Let us temporarily sidestep this decree and observe that
\begin{equation}
-\left(\frac{{\rm d}^2}{{\rm d}y^2}+y^2\right){\rm e}^{\pm{\rm i}y^2/2}=\mp{\rm i}\,{\rm e}^{\pm{\rm i}y^2/2}
\label{dekret}
\end{equation}
is very reminiscent of the equation governing the quantum--mechanical vacuum. The thermodynamical density corresponding to the state $\exp(\pm{\rm i}y^2/2)$ equals the constant unit function on $\mathbb{R}$, which is nonnormalisable under $\vert\vert\cdot\vert\vert_1$ in $L^1(\mathbb{R})$ but carries finite norm under $\vert\vert\cdot\vert\vert_{\infty}$ in $L^{\infty}(\mathbb{R})$. As a perfectly uniform probability distribution, {\it $\exp(\pm{\rm i}y^2/2)$ is the thermodynamical state that maximises the entropy}\/. All the eigenstates in (\ref{aniv}) are thermodynamical excitations thereof, hence they carry less entropy. Of course, we cannot allow the eigenvalues $\sigma=\pm{\rm i}$ within our spectrum, but the above discussion is illustrative because, by (\ref{asstt}),  all our thermodynamical eigenstates (\ref{aniv}) tend asymptotically to a linear combination of the states $y^{-1/2}\exp\left[\pm\frac{{\rm i}}{2}\left(\sigma\ln(y)+y^2\right)\right]$. In other words, all our thermodynamical eigenstates can be interpreted as {\it fluctuations around a state of maximal entropy}\/.

\subsection{Irreversibility {\it vs}\/. nonunitarity}\label{irrvsnon}

A key consequence of irreversibility is nonunitarity. Contrary to the operators $U(t)$ of (\ref{uniprop}), which are unitary on $L^2(\mathbb{R})$, the operators ${\cal U}(\tau)$  of (\ref{juguete}) are {\it non}\/unitary on $L^{\infty}(\mathbb{R})$.

Nonunitarity is readily proved. Let $w_{\sigma}\in L^{\infty}(\mathbb{R})$ be such that ${\cal H}_{\rm out}w_{\sigma}=\sigma w_{\sigma}$. Since $\sigma\in\mathbb{R}$ we have, by (\ref{juguete}),
\begin{equation}
{\cal U}(\tau)w_{\sigma}=\exp\left(-\frac{\tau\sigma}{2k_B}\right)w_{\sigma}, \qquad \tau\sigma\in\mathbb{R},
\label{entoncew}
\end{equation}
hence
\begin{equation}
\vert\vert {\cal U}(\tau)w_{\sigma}\vert\vert_{\infty} = \exp\left(-\frac{\tau\sigma}{2k_B}\right)\vert\vert w_{\sigma}\vert\vert_{\infty}, \qquad \tau\sigma\in\mathbb{R},
\label{nnor}
\end{equation}
which proves our assertion. To summarise: combining (\ref{juguete}), (\ref{miglia}) and (\ref{venti}) we find, after reinstating dimensional factors, that the operators 
\begin{equation}
{\cal U}(\tau)=\exp\left[\frac{\tau}{2k_B}\left(\frac{1}{2R}\frac{{\rm d}^2}{{\rm d}y^2}+\frac{R\gamma^2}{2}y^2\right)\right], \qquad \tau\geq 0,
\label{uopp}
\end{equation}
provide a {\it non}\/unitary, infinite--dimensional representation of the Chapman--Kolmogorov semigroup  (\ref{ohlala}) on $L^{\infty}(\mathbb{R})$. The space $L^1(\mathbb{R})$ also carries an infinite--dimensional representation of (\ref{ohlala}) on which the operators (\ref{uopp}) act.

It is interesting to observe that the eigenfunctions in (\ref{dekret}), which we have discarded for reasons already explained, circumvent the above proof because their eigenvalues are purely imaginary. Each one of them actually provides a 1--dimensional, {\it unitary}\/ representation of (\ref{ohlala}) on $L^{\infty}(\mathbb{R})$.

\section{Discussion}\label{disku}

Classical thermodynamics is the paradigm of emergent theories. It renounces the detailed knowledge of a large number of microscopic degrees of freedom, in favour of a small number of macroscopic averages that retain only some coarse--grained features of the system under consideration. It has been claimed in the literature that quantum mechanics must be an emergent theory \cite{ADLER, CARROLL, THOOFT3}. As one further piece of evidence in support of this latter statement, in this paper we have developed a thermodynamical formalism for quantum mechanics. 

In the usual formulation of quantum theory, one is concerned with the matrix elements $\langle\psi\vert{\cal O}\vert\phi\rangle$ of some operator ${\cal O}$, where the incoming state $\vert\phi\rangle$ belongs to $L^2(\mathbb{R})$ and the outgoing state $\langle\psi\vert$ belongs to the topological dual space, again $L^2(\mathbb{R})$.

In the thermodynamical theory that is dual to quantum mechanics one is again concerned with matrix elements of the type 
$(\psi\vert{\cal O}\vert\phi)$. However, now the incoming state is not square integrable but just integrable, $\vert\phi)\in L^1(\mathbb{R})$, while the outgoing state $(\psi\vert\in L^{\infty}(\mathbb{R})$  belongs to a totally different space. Neither $L^1(\mathbb{R})$ nor its topological dual $L^{\infty}(\mathbb{R})$ qualify as a Hilbert space, because their respective norms do not derive from a scalar product; they are just Banach spaces. The absence of a scalar product is the hallmark of irreversibility. Indeed the thermodynamics that is dual to quantum mechanics is that of irreversible processes (considered here in the linear regime).

One is often interested in the case when the operator ${\cal O}$ is the time evolution operator ${\cal U}$ connecting the incoming and the outgoing states.  Not being allowed to exchange the incoming and the outgoing {\it states}\/ in the transition probability $(\psi\vert{\cal U}\vert\phi)$, because they belong to different spaces, emphasis falls on the {\it process}\/ ${\cal U}$ connecting these two. Irreversibility manifests itself through the nonunitarity of the representation constructed here for the Chapman--Kolmogorov equation. The latter is the functional equation satisfied by ${\cal U}$.

Incoming states $\vert\phi)\in L^1(\mathbb{R})$ are probability densities, as opposed to the probability density {\it amplitudes}\/ $\vert\phi\rangle\in L^2(\mathbb{R})$ of standard quantum theory. Outgoing states $(\psi\vert\in L^{\infty}(\mathbb{R})$ have a different physical interpretation. The norm $\vert\vert\cdot\vert\vert_{\infty}$ can be regarded as a probability density that is {\it not}\/ meant to be integrated. Indeed a general function $\psi\in L^{\infty}(\mathbb{R})$ need not be normalisable under the norms $\vert\vert\cdot\vert\vert_1$ and $\vert\vert\cdot\vert\vert_2$ on $L^1(\mathbb{R})$ and $L^2(\mathbb{R})$ respectively. There is nothing unusual about this---scattering states in standard quantum theory also give rise to nonnormalisable probability densities. 

As an example, in section \ref{banij} we have worked out the spectrum for the thermodynamical harmonic oscillator. This implies solving the Schroedinger equation for the {\it repulsive}\/ potential $V(y)=-y^2$, the wrong sign being due to the Wick rotation connecting irreversible thermodynamics to mechanics. Not surprisingly, the spectrum is empty when diagonalising the Hamiltonian on the space $L^1(\mathbb{R})$, while exhibiting rich features on the space $L^{\infty}(\mathbb{R})$. In particular, all our eigenstates turn out to be nonnormalisable under the norms $\vert\vert\cdot\vert\vert_1$ and $\vert\vert\cdot\vert\vert_2$ on $L^1(\mathbb{R})$ and $L^2(\mathbb{R})$ respectively, hence they all are analogous to scattering states in standard quantum theory. However all our eigenstates are normalisable under the norm $\vert\vert\cdot\vert\vert_{\infty}$ of $L^{\infty}(\mathbb{R})$. 

An apparently striking feature is the reluctance of incoming states to build {\it eigenstates}\/ of the Hamiltonian, as seen in section \ref{spek}. This apparent difficulty disappears once one realises that {\it outgoing}\/ states make perfectly good eigenstates.  Furthermore,  the existence of outgoing states that cannot be reached by the time evolution of any incoming state whatsoever is another sign of irreversibility. We cannot renounce irreversibility because we have programatically dispensed with time reversal symmetry. Hence incoming eigenstates must go.

\vskip.5cm
\noindent
{\it Exeunt omnes.}

\end{document}